# Directional emission from WS$_2$ monolayer coupled to plasmonic Nanowire-on-Mirror Cavity

*Shailendra K. Chaubey[1], Gokul M. A.[1], Diptabrata Paul[1], Sunny Tiwari[1], Atikur Rahman[1], G.V. Pavan Kumar[1,*]*

[1]Department of Physics, Indian Institute of Science Education and Research, Pune-411008, India

*E-mail: pavan@iiserpune.ac.in

Keywords: Fourier plane Imaging, WS$_2$, 2D Material, Trion, Exciton, Plasmonics, Directional Emission

Abstract:

Influencing spectral and directional features of exciton emission characteristics from 2D transition metal dichalcogenides by coupling it to plasmonic nano-cavities has emerged as an important prospect in nanophotonics of 2D materials. In this paper we experimentally study the directional photoluminescence emission from Tungsten disulfide (WS$_2$) monolayer sandwiched between a single-crystalline plasmonic silver nanowire (AgNW) waveguide and a gold (Au) mirror, thus forming an AgNW-WS$_2$-Au cavity. By employing polarization-resolved Fourier plane optical microscopy, we quantify the directional emission characteristics from the distal end of the AgNW-WS$_2$-Au cavity. Given that our geometry simultaneously facilitates local field enhancement and waveguiding capability, we envisage its utility in 2D material-based, on-chip nanophotonic signal processing, including nonlinear and quantum optical regimes.

## 1. Introduction:

In the last few years, 2D transition metal dichalcogenides (TMDs) have attracted significant attention because of their novel optical and electronic properties.[1-4] Direct bandgap in their monolayer makes them a suitable candidate for electronics applications[3-5]. Very high oscillator strength leads to a narrow peak with a pronounced optical transition.[6,7] In addition, TMDs exhibit valley polarization effects, making them suitable candidate for spin-orbit interaction studies.[8-10] Coupling 2D materials such as tungsten disulfide ($WS_2$) monolayer to plasmonic nano-cavities can enhance and influence light-matter coupling. In this paper, we report on our experimental studies of directional photoluminescence from $WS_2$ monolayer sandwiched in a unique kind of cavity: Ag nanowire on a gold mirror.

Plasmonic nano-cavities are important architectures to study because of the very high electric field inside the cavity caused by hybridization of gap plasmon.[11] Ultrasmall volume inside the cavity provides a large Purcell enhancement factor.[12,13] To this end, nano-cavity formed by a nanoparticle on gold film has been utilized for strong-coupling at the single-molecule level,[14] single-molecule Surface-Enhanced Raman Scattering (SERS),[15,16] controlled reflectance properties,[17] enhanced spontaneous emission.[13,18]

Of relevance to this study is the Ag Nanowire on a gold mirror cavity. Nanocavity formed by a silver nanowire (AgNW) on gold film, apart from the above advantages, can also work as an opto-plasmonic waveguide of cavity's emission, as shown recently by our group.[19] Such nanocavities are also studied in the context of Purcell enhancement,[18] remote excitation of the molecules,[19] spontaneous emission enhancement,[18] wave-vector distribution[19] and trion enhancement.[20]

Coupling surface plasmon polaritons to TMDs can facilitate interesting optical properties. To this end, studying the optical transition characteristics of TMDs in vicinity of a plasmonic nanostructure near resonance has gained relevance.[21-24] A variety of prospects like

Photoluminescence (PL) and Raman enhancement,[25,26] enhanced spin-orbit interaction,[20,27] remote excitation of SERS,[28] spectrum tailoring,[29] strong coupling[30] and trion enhancement[20] has been achieved using such configuration. For these purposes many plasmonic structures like bowtie antenna,[31] nano-disk array,[32] nano-cube,[29] nanoparticle[30] and nanowire[20, 28] has been either fabricated over TMDs[33] or TMDs is transferred onto the structures.[34] Specifically, in the context of 2D materials, TMDs have been coupled to a single AgNW for studying remote SERS,[27] second harmonic generation,[35-38] logic operation,[39] Rabi splitting[40] and plasmon-exciton interconversion[20]. Silver film - Ag nanowire cavity has been recently utilized for the Trion enhancement and enhancing the spin-orbit coupling.[20]

Apart from these effects, a plasmonic Ag NW can act as a subwavelength waveguide as well as a nanoscale antenna which can be harnessed to route an excited signal and emit the signal at the distal end.[41] This utility of Ag nanowire both as a plasmonic cavity and as a waveguide for 2D materials has not been explored, which we do in this paper.

**Figure 1** shows the studied geometry. It contains a $WS_2$ monolayer sandwiched between a plasmonic silver nanowire and a gold mirror (AgNW-$WS_2$-Au cavity). A thin, $Al_2O_3$ layer is deposited on gold mirror which acts as a buffer layer. In such waveguide-cavity systems, wavevector analysis becomes an important aspect of study to quantify the emission process. Motivated by this, our study focuses on the wavevector distribution of the PL emission in AgNW-$WS_2$-Au cavity. Specifically, we excite one end of the cavity with a focused 532nm laser and study the PL spectrum from the other end of the cavity as a function of angle and polarization. To achieve this, we employ polarization-resolved Fourier-plane optical microscopy to study the emission direction of AgNW-$WS_2$-Au cavity and quantify the in-plane angular distribution. By studying the angular distribution and polarization-resolved spectra, we show that AgNW-WS$_2$-Au cavity can modify the spectral feature of WS$_2$ monolayer by interconversion between exciton and trion.

## 2. Experimental Section:

**2.1 WS$_2$ Synthesis:** The WS$_2$ monolayers were grown using atmospheric pressure chemical vapor deposition (APCVD) on 300 nm SiO$_2$ coated silicon wafer following the procedure mentioned in ref. [42] and [43]. The substrates were sonicated in acetone and IPA for 10 min each and blow dried. They were further cleaned using O$_2$ plasma at 60 W for 5 min. The substrates were loaded on to an alumina boat containing 500 mg of WO$_3$ with the smooth side facing the powder. The boat containing the WO$_3$ was placed inside a quartz tube of 3.5 cm inner diameter in the heating zone of the furnace. Another boat containing 500 mg of Sulphur was placed upstream inside the tube, 15 cm away from the WO$_3$ boat. The Sulphur boat was placed outside the furnace and was heated separately using a heater coil. Initially, the tube was flushed with 500 standard cubic centimeter per minute (SCCM) of argon for 10 min, and then the flow was reduced to 30 SCCM and was maintained throughout the experiment. The furnace was ramped up to 850 $^0$C at a rate of 5 $^0$C/min. As the furnace reached 850 $^0$C the heater coil was heated to 240 $^0$C to evaporate Sulphur. These temperatures were maintained for 10 min for growing WS$_2$ monolayer. After the growth, the system was allowed to cool naturally.

**2.2 Optical Characterization:** Chemically synthesized WS$_2$ monolayer was further characterized by PL and Raman spectroscopy. PL and Raman spectra of the same is shown in the supplementary information S1.

**2.3 AgNW-WS$_2$-Au cavity preparation:** Ag nanowires with average diameter 300 nm have been synthesized by polyol process as reported in.[44] For the preparation of the gold mirror, 160 nm gold film has been deposited on the glass coverslip using thermal vapor deposition. Monolayer WS$_2$ flakes have been transferred onto the gold mirror. Polystyrene is used as a support film for this transfer as reported here.[45] 3nm Al$_2$O$_3$ spacer layer was placed using atomic layer deposition in between gold mirror and WS$_2$ to prevent the charge screening with the aim to avoid PL quenching. To form a cavity, Ag nanowires were dropcasted on gold mirror with WS$_2$ sandwiched in between them. NW at which we have performed the experiment is 300 nm thick and approximately 13 µm in length. For different set of measurements shown in supplementary information, NWs of length 12-20 µm have been used. FESEM image showing the diameter of the wire is given in supplementary information S7.

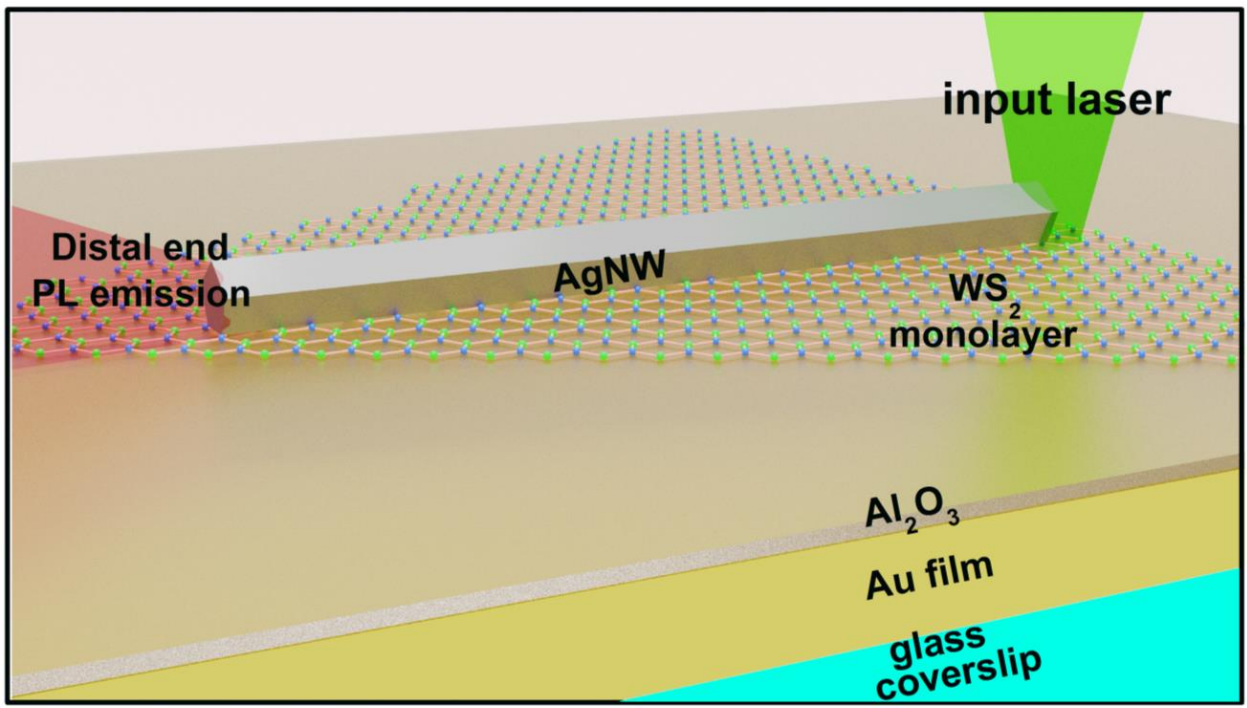

**Figure 1.** Schematic of the experimental setup. WS$_2$ monolayer was placed over a gold film separated by a 3 nm Al$_2$O$_3$ spacer layer. Single Ag NW was placed over the monolayer WS$_2$ flake. One end of the nanowire

was excited with 532 nm laser. WS$_2$ PL from the excitation point couples to the nanowire plasmons, which is further out-coupled from the distal end of the nanowire. The emission from the distal end was collected and was projected to spectrometer and EMCCD for spectroscopy and Fourier plane imaging, respectively.

**2.4 Experimental Setup:** One end of AgNW-WS$_2$-Au cavity has been excited using 532 nm laser with polarization along the cavity. 100x, 0.95 numerical aperture (NA) objective lens was used in backscattered configuration for both excitation and collection. Signal from the distal end of the nanowire is collected by spatially filtering the region, and the real/Fourier plane is projected into EMCCD/spectrometer. To transfer the Fourier plane from the back aperture of the objective lens to the EMCCD, 4f configuration is used.[46] High NA excitation ensures efficient excitation of surface plasmon in the nanowire as well as high electric field in the cavity. Combination of edge and notch filter has been used to efficiently reject the elastically scattered light. It can be seen from the PL spectra that the contribution from the elastic scattering is negligible (supplementary information S1). Polarizer and half wave plate is used in input path to control the input polarization. An analyzer is used in output path to analyze the output light for polarization resolved measurements. See supplementary information S2 and our previous reports[47] for detailed experimental setup of Fourier plane optical microscopy.

**3. Result and Discussion:**

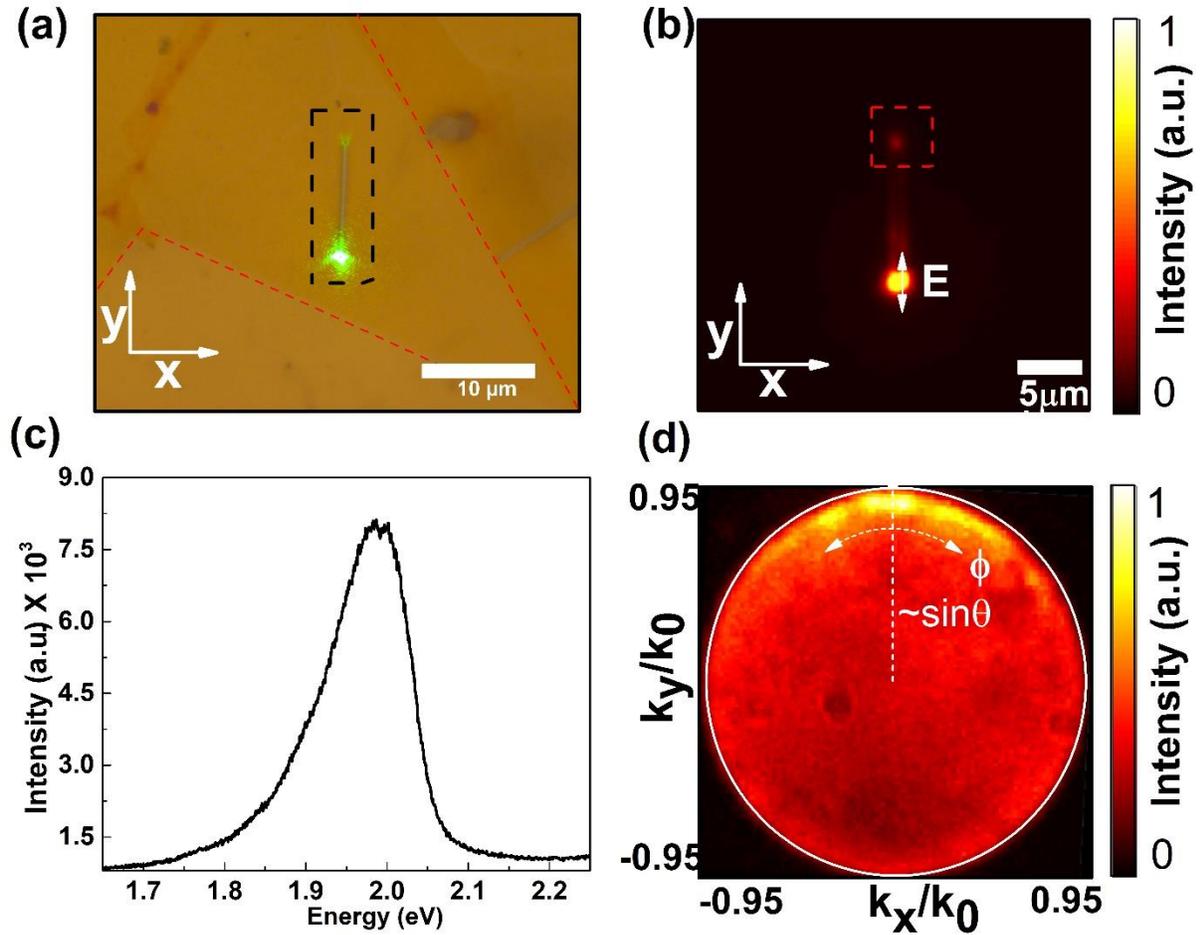

**Figure 2.** Directional emission from the distal end of Ag nanowire (a) Bright-field image of nanowire over $WS_2$, which is placed over a gold mirror with 3 nm spacer layer of $Al_2O_3$. $WS_2$ monolayer boundaries are shown with red dotted lines. One end of the nanowire was excited with 532 nm laser with polarization along the long axis of the nanowire. (b) PL image of the nanowire after rejecting the elastically scattered light using combination of edge and notch filters. Dotted box shows the distal end collection of the nanowire. (c) $WS_2$ PL spectrum collected from the distal end of the nanowire as per the configuration in (b). (d) Fourier space intensity distribution of the PL out-coupled from the distal end of the nanowire showing directional emission in narrow range of wavevectors.

**3.1 Directional photoluminescence from AgNW-WS$_2$-Au cavity: Figure 2** a is the bright-field image of the Ag nanowire placed on a gold mirror AgNW-WS$_2$-Au cavity. Excitation of nanowire end, with a tightly focused 532 nm laser excites the propagating plasmons along the nanowire. In addition, the excitation of hybridized gap plasmons between the AgNW and metal film, creates high local electric field, which enhances the PL emission from $WS_2$ monolayer. Because of the near field interaction, the PL emission gets coupled to the nanowire SPPs travelling along the length of the nanowire. Because of the spatial discontinuity at the nanowire end, these SPPs are out-coupled as free space photons. In figure 2b, we observe strong PL emission from AgNW-$WS_2$-Au cavity not only from the excitation and distal ends, but also throughout the nanowire. This is because of the high electric field in the nanowire on a metal film cavity, which acts as a hot-line along the length of the cavity. Figure 2c shows the emission spectrum collected from the distal end of AgNW-$WS_2$-Au cavity. To study the wavevector of emission from the cavity, we performed Fourier plane imaging, which maps the emission wavevectors in terms of θ and φ spreading. Radial coordinate in Fourier plane is $NA = n\,sin\theta$, and φ is tangential coordinate varies from 0 to 2π. Fourier plane image Figure 2d shows that the emission from the nanowire end is directed towards higher $k_y/k_0$ and covers only a small range of radial and azimuthal angles, indicating highly directional emission. Multiple measurements for the momentum space image is shown in Supplementary information S4.

A relevant question to ask is how does the presence of gold mirror influence the performance of the cavity on and off the absorption band of $WS_2$ monolayer? To explore this, we performed experiments similar to figure 2, but placing the $WS_2$ on a glass coverslip and dropcasting the Ag nanowire over it and exciting it with 532nm and 633 nm lasers. In the case of 532 nm excitation, we do not observe any plasmon-assisted propagation from the distal end of the nanowire because of high absorption of incoming laser by $WS_2$ and NW, whereas for 633 nm excitation, we do observe plasmon propagation (see supplementary information S3). Although we observe plasmon propagation with 633 nm, the PL of WS$_2$ cannot be excited at this wavelength. To understand the role of AgNW-Au cavity we have calculated the modes of the cavity. For the AgNW-Au cavity, we have observed eight modes. The higher order SPP mode (see supplementary information S6 (a) and (b)) have larger propagation length as their imaginary part of effective mode index is small. In contrast, six bound modes (see supplementary information S6 (c) - (h)) exhibit high loss owing to very high imaginary component of the mode index. These higher order mode for which near field is distributed near the top of the NW contributes to the propagation.[48] An inference we draw from this is that the presence of gold mirror enhances the propagation length.

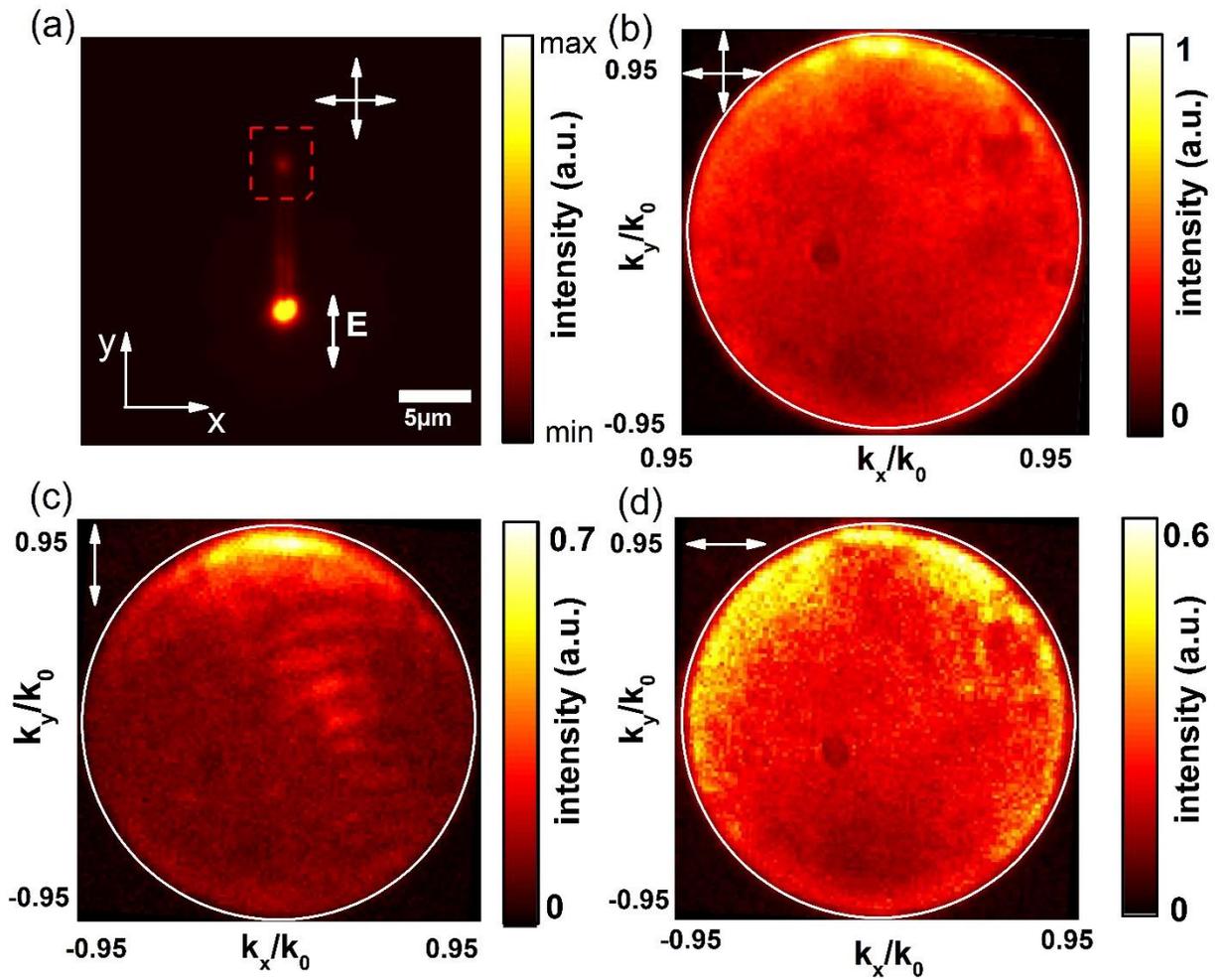

**Figure 3.** Emission-polarization-resolved Fourier plane imaging of distal end PL emission, which indicates the momentum-space distribution of the emitted light from a particular region in the geometry. (a) Real-plane PL intensity distribution of the nanowire captured using EMCCD integrated over, all polarizations (indicated by orthogonal arrows); (b) corresponding momentum space PL intensity distribution, collected from the distal end of the nanowire (shown by dotted box in (a)). Momentum space PL intensity distribution of the emission from the distal end of the nanowire when polarization is analyzed (c) along the long axis of the nanowire and (d) transverse to the long axis of the nanowire.

**3.2 Polarization resolved angular emission from AgNW-WS$_2$-Au cavity:** The emission at the distal end of nanowire is mediated through the SPPs of the nanowire, which generally shows rich polarization signature. In addition to this, emission from the WS$_2$ which is confined in the AgNW-WS$_2$-Au cavity is enhanced and out-couples through the nanowire end, after propagating through the cavity. To study the effect of nanowire and metal film cavity, we performed polarization resolved Fourier plane imaging on the PL emission from the distal end of nanowire. **Figure 3** a represents the PL image of the AgNW-WS$_2$-Au cavity after rejecting the elastically scattered light captured using EMCCD. Figure 3b is the momentum space image of the PL emission from the distal end of the nanowire. Figure 3c is the momentum space image analyzed for the polarization along the long axis of the nanowire and Figure 3d represents the momentum space image analyzed for the polarization transverse to nanowire. For the output polarization along the nanowire, we observe that light is directional and most of the emission is centered around $\varphi = 0$. While in case of the output polarization transverse to the axis of the nanowire, the $\varphi$ spread is more. This means that photons which are polarized along the wire are more directional in comparison to the photon whose polarization is transverse to the nanowire. This is attributed to the polarization maintaining properties of surface plasmon polaritons in a nanowire.[49]

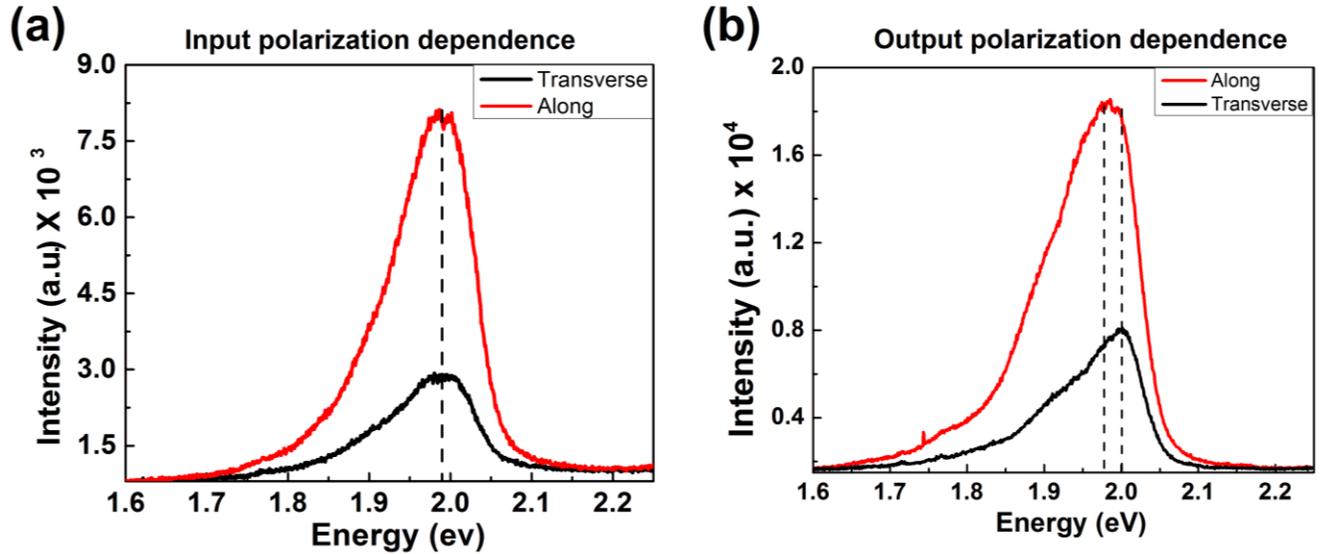

**Figure 4.** Dependence of PL spectral features from the distal end of AgNW-WS$_2$-Au cavity as a function of excitation and collection polarization. (a) PL spectrum as a function of excitation polarization along (red curve) and perpendicular (black curve) to the nanowire axis. (b) PL spectrum as a function of analyzed polarization along (red curve) and perpendicular (black curve) to the nanowire axis. Apart from the reduction in the intensity, the change in polarization also leads to redshift in the peaks.PL spectrum collected from the distal end of the nanowire, red and black curve corresponds to the input polarization along the long axis of the nanowire, and transverse to the long axis of the nanowire.

**3.3 Polarization-resolved PL spectral characteristics in AgNW-WS$_2$-Au cavity**: In previous section, we discussed about how polarized excitation and collection can influence the directional emission from AgNW-WS$_2$-Au cavity. This further motivated us to study the spectral characteristics as a function of excitation and collection polarization. First we checked the excitation polarization dependence. The red and black curves in **Figure 4**a are the PL spectrum from the distal end of the nanowire AgNW-WS$_2$-Au cavity when input excitation polarization is along and transverse to the long axis of the nanowire, respectively. It can be clearly seen that output PL intensity is greater when excitation polarization is along the long axis of the nanowire. With this hindsight, we fixed the excitation polarization along the long axis of the wire, and collected the polarization resolved PL emission from distal end. In figure 4b, the red and black curve represents the PL emission along and perpendicular to the long axis of the nanowire. It can be clearly seen that intensity in the case of output polarization along the long axis of the nanowire is more in comparison to the transverse output polarization. It is also observed that the PL peak in case of the output polarization along the long axis of the nanowire is red-shifted in comparison to the output polarization transverse to the long axis of the nanowire. This redshift is found to be approximately $\Delta E = 23$ meV.

Since the PL spectrum has the contribution of both the neutral exciton and the charged exciton (trion), such redshift can be attributed to the exciton to trion conversion.[50] Trions are essentially formed when exciton is bound by an excess of electron or hole, also called charged exciton. Photoionization and doping are the two main way to convert exciton into trion [51,52]. In this case, WS$_2$ is slightly n doped, so the trions in the system are negatively charged. Exciton to trion conversion is reported by coupling with SPP in MIM cavity (metal insulator metal) cavity. [20] In AgNW-WS$_2$-Au cavity there are two kind of plasmon involved. One is propagating along the wire and second is localized in the cavity.[53] The intensity component along the NW is larger as compared to the transverse to the NW.

At higher irradiation power, it has been observed that exciton converts into trion; which leads to redshift in PL spectra.[52] To further confirm this, we deconvoluted the PL spectra into two peaks by Voigt function double fit (see S5 of supplementary information). The deconvoluted spectrum is consistent with the reported exciton and Trion peak,[51] and Trion binding energy is found to be around 50 meV, which is consistent with the reported value[54,20]. However, the propagating SPP may experience reabsorption by $WS_2$, NW or gold film during the propagation which leads to the redshift for the component having polarization along the wire. Absorption coefficient is fairly constant above 550 nm for $WS_2$,[55] so the contribution from the reabsorption of $WS_2$ is less probable, while reabsorption by Ag NW or metal film may be a possible reason for the above redshift.[56]

**Conclusion:**

To summarize, we have experimentally studied the polarization resolved directional PL emission from AgNW-$WS_2$-Au cavity as a function of angular and spectral signatures. Through detailed experiments and analysis, we have shown the role of each component of the cavity, and how it influences the directional emission characteristics of the $WS_2$ monolayer. By performing polarization- and momentum-resolved PL spectroscopy from the distal end of the AgNW-$WS_2$-Au cavity, we reveal the possibility of studying exciton to trion conversion at room temperature. It is to be noted that nanowire based cavities inherently facilitate waveguiding characteristics, which means the excitation and collection in the geometry is spatially off-set. Such separation of excitation and collection of optical locations can be of relevance in 2D material based nonlinear nanophotonic chips and open-cavity nano-lasers. We envisage that the confined electric fields facilitated by the spatially-extended cavity can be further harnessed by engineering various parameters such as: composition, shape and size of the nanowire and the underlying gold mirror.

**Supporting Information**

Supporting Information is available from the Wiley Online Library or from the author.

**Acknowledgement:**

This work was partially funded by Air Force Research Laboratory grant (FA2386-18-1-4118 R&D 18IOA118), DST Energy Science grant (SR/NM/TP-13/2016) and Swarnajayanti fellowship grant (DST/SJF/PSA02/2017-18) to GVPK. SKC and GVPK would also like to thank Mandar Deshmukh and Mahesh Gokhale from TIFR, Mumbai for their help in atomic layer deposition. Authors would also thank Chetna Taneja and Vandana Sharma for fruitful discussion.

**References:**

[1]    K. F. Mak, C. Lee, J. Hone, J. Shan, T. F. Heinz, Physical Review Letters 2010, 105, 136805.

[2]    A. Splendiani, L. Sun, Y. Zhang, T. Li, J. Kim, C.-Y. Chim, G. Galli, F. Wang, Nano Letters 2010, 10, 1271.

[3]    Q. H. Wang, K. Kalantar-Zadeh, A. Kis, J. N. Coleman, M. S. Strano, Nature Nanotechnology 2012, 7, 699.

[4]    K. Roy, M. Padmanabhan, S. Goswami, T. P. Sai, G. Ramalingam, S. Raghavan, A. Ghosh, Nature Nanotechnology 2013, 8, 826.

[5]    R. Deshmukh, P. Marques, A. Panda, M. Y. Sfeir, S. R. Forrest, V. M. Menon, ACS Photonics 2020, 7, 43; J. P. Mathew, G. Jegannathan, S. Grover, P. D. Dongare, R. D. Bapat, B. A. Chalke, S. C. Purandare, M. M. Deshmukh, Applied Physics Letters 2014, 105, 223502.

[6] A. Chernikov, T. C. Berkelbach, H. M. Hill, A. Rigosi, Y. Li, O. B. Aslan, D. R. Reichman, M. S. Hybertsen, T. F. Heinz, Physical review letters 2014, 113, 076802.

[7] K. He, N. Kumar, L. Zhao, Z. Wang, K. F. Mak, H. Zhao, J. Shan, Physical review letters 2014, 113, 026803.

[8] D. Xiao, G.-B. Liu, W. Feng, X. Xu, W. Yao, Physical review letters 2012, 108, 196802.

[9] H. Zeng, J. Dai, W. Yao, D. Xiao, X. Cui, Nature Nanotechnology 2012, 7, 490.

[10] "Long Range Valley Hall Effect in WS2 Bloch Surface Wave Exciton Polaritons", presented at *Conference on Lasers and Electro-Optics*, Washington, DC, 2020/05/10, 2020.

[11] E. Prodan, C. Radloff, Science 2003, 302, 419.

[12] P. Lodahl, S. Mahmoodian, S. Stobbe, Reviews of Modern Physics 2015, 87, 347.

[13] G. M. Akselrod, C. Argyropoulos, T. B. Hoang, C. Ciracì, C. Fang, J. Huang, D. R. Smith, M. H. Mikkelsen, Nature Photonics 2014, 8, 835.

[14] R. Chikkaraddy, B. de Nijs, F. Benz, S. J. Barrow, O. A. Scherman, E. Rosta, A. Demetriadou, P. Fox, O. Hess, J. J. Baumberg, Nature 2016, 535, 127.

[15] E. C. Le Ru, P. G. Etchegoin, in *Principles of Surface-Enhanced Raman Spectroscopy*, (Eds: E. C. Le Ru, P. G. Etchegoin), Elsevier, Amsterdam 2009, 1.

[16] G. AlaguVibisha, J. K. Nayak, P. Maheswari, N. Priyadharsini, A. Nisha, Z. Jaroszewicz, K. B. Rajesh, R. Jha, Optics Communications 2020, 463, 125337.

[17] A. Moreau, C. Ciracì, J. J. Mock, R. T. Hill, Q. Wang, B. J. Wiley, A. Chilkoti, D. R. Smith, Nature 2012, 492, 86.

[18] K. J. Russell, T.-L. Liu, S. Cui, E. L. Hu, Nature Photonics 2012, 6, 459.

[19] A. B. Vasista, H. Jog, T. Heilpern, M. E. Sykes, S. Tiwari, D. K. Sharma, S. K. Chaubey, G. P. Wiederrecht, S. K. Gray, G. V. P. Kumar, Nano Letters 2018, 18, 650.

[20] J. Shi, J. Zhu, X. Wu, B. Zheng, J. Chen, X. Sui, S. Zhang, J. Shi, W. Du, Y. Zhong, Q. Wang, Q. Zhang, A. Pan, X. Liu, Advanced Optical Materials 2020, 8, 2001147.

[21] P. Sriram, A. Manikandan, F.-C. Chuang, Y.-L. Chueh, Small 2020, 16, 1904271.

[22] J. Gu, B. Chakraborty, M. Khatoniar, V. M. Menon, Nature Nanotechnology 2019, 14, 1024.

[23] P. Ni, A. De Luna Bugallo, V. M. Arellano Arreola, M. F. Salazar, E. Strupiechonski, V. Brändli, R. Sawant, B. Alloing, P. Genevet, ACS Photonics 2019, 6, 1594.

[24] P. Ni, A. De Luna Bugallo, X. Yang, V. M. Arellano Arreola, M. Flores Salazar, E. Strupiechonski, B. Alloing, C. Shan, P. Genevet, Journal of Physics D: Applied Physics 2019, 52, 374001.

[25] F. Cheng, A. D. Johnson, Y. Tsai, P.-H. Su, S. Hu, J. G. Ekerdt, C.-K. Shih, ACS Photonics 2017, 4, 1421.

[26] H. S. Lee, M. S. Kim, Y. Jin, G. H. Han, Y. H. Lee, J. Kim, Physical Review Letters 2015, 115, 226801.

[27] Q. Guo, T. Fu, J. Tang, D. Pan, S. Zhang, H. Xu, Physical Review Letters 2019, 123, 183903.

[28] N. Coca-López, N. F. Hartmann, T. Mancabelli, J. Kraus, S. Günther, A. Comin, A. Hartschuh, Nanoscale 2018, 10, 10498.

[29] J. Huang, G. M. Akselrod, T. Ming, J. Kong, M. H. Mikkelsen, ACS Photonics 2018, 5, 552.

[30] M.-E. Kleemann, R. Chikkaraddy, E. M. Alexeev, D. Kos, C. Carnegie, W. Deacon, A. C. de Pury, C. Große, B. de Nijs, J. Mertens, A. I. Tartakovskii, J. J. Baumberg, Nature Communications 2017, 8, 1296.

[31]    E. Palacios, S. Park, S. Butun, L. Lauhon, K. Aydin, Applied Physics Letters 2017, 111, 031101.

[32]    S. Butun, S. Tongay, K. Aydin, Nano letters 2015, 15, 2700.

[33]    A. D. Johnson, F. Cheng, Y. Tsai, C.-K. Shih, Nano Letters 2017, 17, 4317.

[34]    M. S. Eggleston, S. B. Desai, K. Messer, S. A. Fortuna, S. Madhvapathy, J. Xiao, X. Zhang, E. Yablonovitch, A. Javey, M. C. Wu, ACS Photonics 2018, 5, 2701.

[35]    E. Drobnyh, M. Sukharev, The Journal of Chemical Physics 2020, 152, 094706.

[36]    W.-P. Guo, W.-Y. Liang, C.-W. Cheng, W.-L. Wu, Y.-T. Wang, Q. Sun, S. Zu, H. Misawa, P.-J. Cheng, S.-W. Chang, H. Ahn, M.-T. Lin, S. Gwo, Nano Letters 2020, 20, 2857.

[37]    X. Han, K. Wang, P. D. Persaud, X. Xing, W. Liu, H. Long, F. Li, B. Wang, M. R. Singh, P. Lu, ACS Photonics 2020, 7, 562.

[38]    Y. Li, M. Kang, J. Shi, K. Wu, S. Zhang, H. Xu, Nano Letters 2017, 17, 7803.

[39]    J. H. Kim, J. Lee, H. Kim, S. J. Yun, J. Kim, H. S. Lee, Y. H. Lee, Scientific Reports 2019, 9, 9164.

[40]    D. G. Baranov, B. Munkhbat, E. Zhukova, A. Bisht, A. Canales, B. Rousseaux, G. Johansson, T. J. Antosiewicz, T. Shegai, Nature Communications 2020, 11, 2715.

[41]    T. Shegai, V. D. Miljković, K. Bao, H. Xu, P. Nordlander, P. Johansson, M. Käll, Nano Letters 2011, 11, 706.

[42]    Y. Zhang, Y. Zhang, Q. Ji, J. Ju, H. Yuan, J. Shi, T. Gao, D. Ma, M. Liu, Y. Chen, X. Song, H. Y. Hwang, Y. Cui, Z. Liu, ACS Nano 2013, 7, 8963.

[43]    C. Cong, J. Shang, X. Wu, B. Cao, N. Peimyoo, C. Qiu, L. Sun, T. Yu, Advanced Optical Materials 2014, 2, 131.

[44]    Y. Sun, B. Mayers, T. Herricks, Y. Xia, Nano Letters 2003, 3, 955.

[45]  Z.-Q. Xu, Y. Zhang, S. Lin, C. Zheng, Y. L. Zhong, X. Xia, Z. Li, P. J. Sophia, M. S. Fuhrer, Y.-B. Cheng, Q. Bao, ACS Nano 2015, 9, 6178.

[46]  A. B. Vasista, D. K. Sharma, G. V. P. Kumar, digital Encyclopedia of Applied Physics 2019, 1.

[47]  A. B. Vasista, S. K. Chaubey, D. J. Gosztola, G. P. Wiederrecht, S. K. Gray, G. V. P. Kumar, Advanced Optical Materials 2019, 7, 1900304.

[48]  S. Zhang, H. Xu, ACS Nano 2012, 6, 8128.

[49]  Z. Jia, H. Wei, D. Pan, H. Xu, Nanoscale 2016, 8, 20118.

[50]  P. Fan, B. Zheng, X. Sun, W. Zheng, Z. Xu, C. Ge, Y. Liu, X. Zhuang, D. Li, X. Wang, X. Zhu, Y. Jiang, A. Pan, The Journal of Physical Chemistry Letters 2019, 10, 3763.

[51]  J. J. Carmiggelt, M. Borst, T. van der Sar, Scientific Reports 2020, 10, 17389.

[52]  Y. Kwon, K. Kim, W. Kim, S. Ryu, H. Cheong, Current Applied Physics 2018, 18, 941.

[53]  W. Chen, S. Zhang, Q. Deng, H. Xu, Nature Communications 2018, 9, 801.

[54]  A. Arora, T. Deilmann, T. Reichenauer, J. Kern, S. Michaelis de Vasconcellos, M. Rohlfing, R. Bratschitsch, Physical Review Letters 2019, 123, 167401.

[55]  X. Jiang, B. Sun, Y. Song, M. Dou, J. Ji, F. Wang, RSC Advances 2017, 7, 49309.

[56]  T. Shegai, Y. Huang, H. Xu, M. Käll, Applied Physics Letters 2010, 96, 103114.

# Supplementary Information

# Directional emission from WS$_2$ monolayer coupled to plasmonic Nanowire Cavity

*Shailendra K. Chaubey[1], Gokul M. A.[1], Diptabrata Paul[1], Sunny Tiwari[1], Atikur Rahman[1], G.V. Pavan Kumar[1,*]*

[1] Department of Physics, Indian Institute of Science Education and Research, Pune-411008, India

*E-mail: pavan@iiserpune.ac.in

**1. WS$_2$ characterization before transferring the sample**

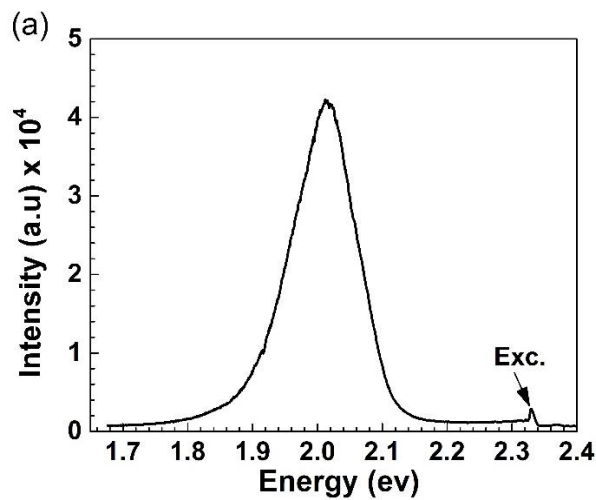 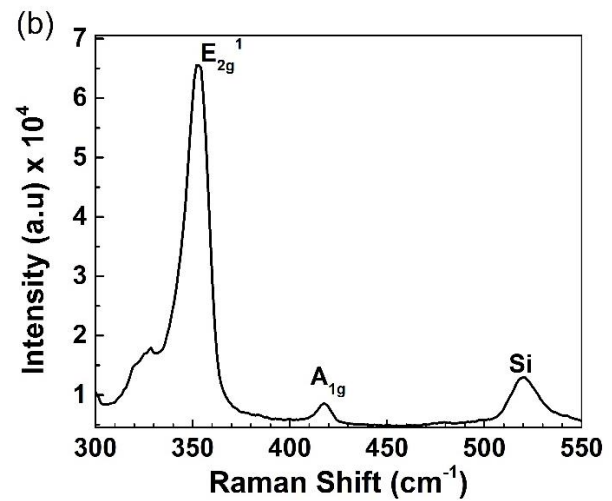

**Figure S1.** Raman and Photoluminescence spectrum of the Monolayer $WS_2$, (a) Photoluminescence spectrum including excitation wavelength to make sure that we are efficiently rejecting the Raleigh scattered light. Excitation wavelength is shown with arrow has negligible intensity in comparison to photoluminescence intensity. (b) Raman spectrum of the monolayer $WS_2$ where peak $E_{2g}^1$ and $A_{2g}$ are at 350 cm$^{-1}$ and 417 cm$^{-1}$. Since $WS_2$ was first grown on Si/SiO$_2$ we can see the silicon peak at 520 cm$^{-1}$.

Photoluminescence (PL) and Raman spectroscopy is performed using 532 nm laser excitation. We see very high PL count which shows that the $WS_2$ on which we are performing the experiment is monolayer. In PL spectrum we have also the excitation wavelength. Small count at excitation wavelength shows that we are effectively rejecting the elastic scattered light. Further Raman measurement shows a large $E_{2g}^1$ at 350 cm$^{-1}$ and small $A_{2g}$ peak at 350 cm$^{-1}$. Intensity of $E_{2g}^1$ peak is an order of magnitude higher than of $A_{2g}$, which further confirms the sample is monolayer $WS_2$. Sample are grown on Si/SiO$_2$ substrate we can see silicon Raman at 520 cm$^{-1}$.

## 2. Experimental Setup

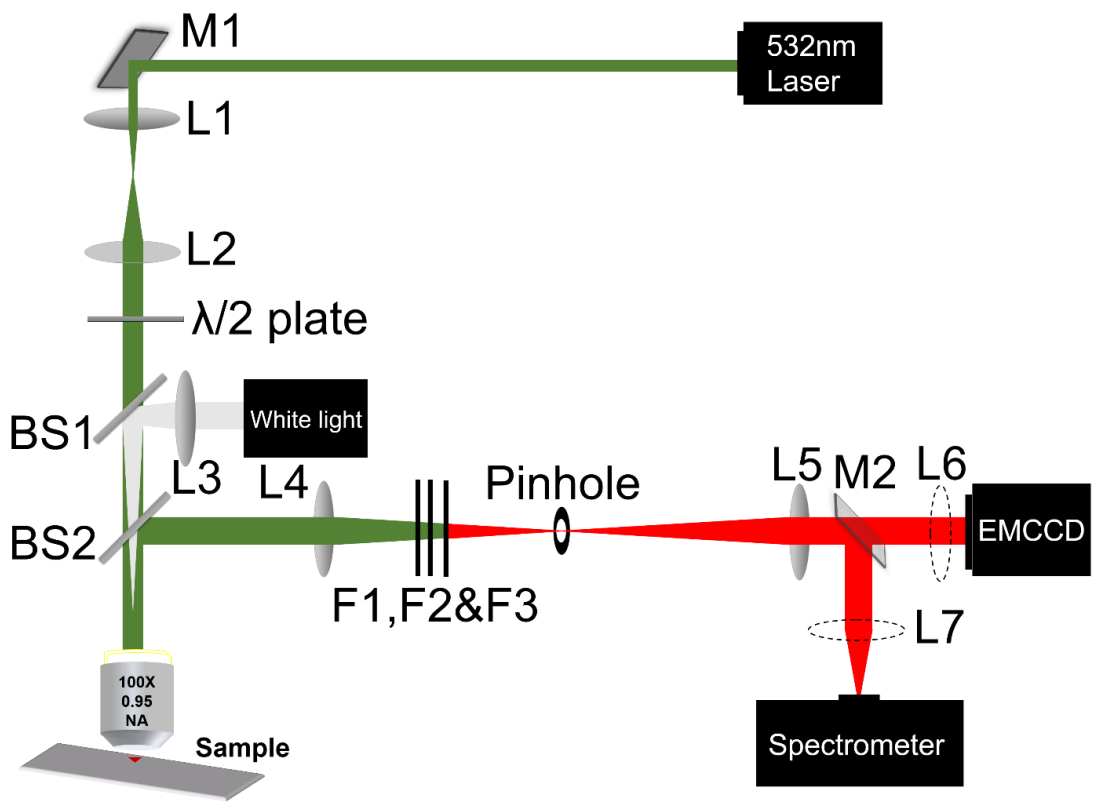

**Figure S2. Experimental setup.**

The sample was excited using a high numerical aperture 100x, 0.95 NA objective lens. The backscattered light was collected using the same lens. The 532 nm laser light was expanded using a set of two lenses L1 and L2. M1 is a mirror. The polarization of the incoming laser was controlled by a λ/2 waveplate in the path. BS1 and BS2 are beam splitters to simultaneously excite the sample with laser and its visualization using white light. Lens L3 is used to loosely focus white light on the sample plane. F1, F2, and F3 are set of two edge filters and one notch filter to reject the elastically scattered light for SERS spectroscopy and Fourier plane and energy-momentum imaging. Lenses L4 and L5 are used to project the emission to the Fourier plane onto the spectrometer or EMCCD. M2 is a flip mirror, used to project the light on the spectrometer for spectroscopy and energy-momentum imaging. Lenses L6 and L7 are flip lenses used to switch from real plane to Fourier plane.

## 3. Wire and WS$_2$ over the glass

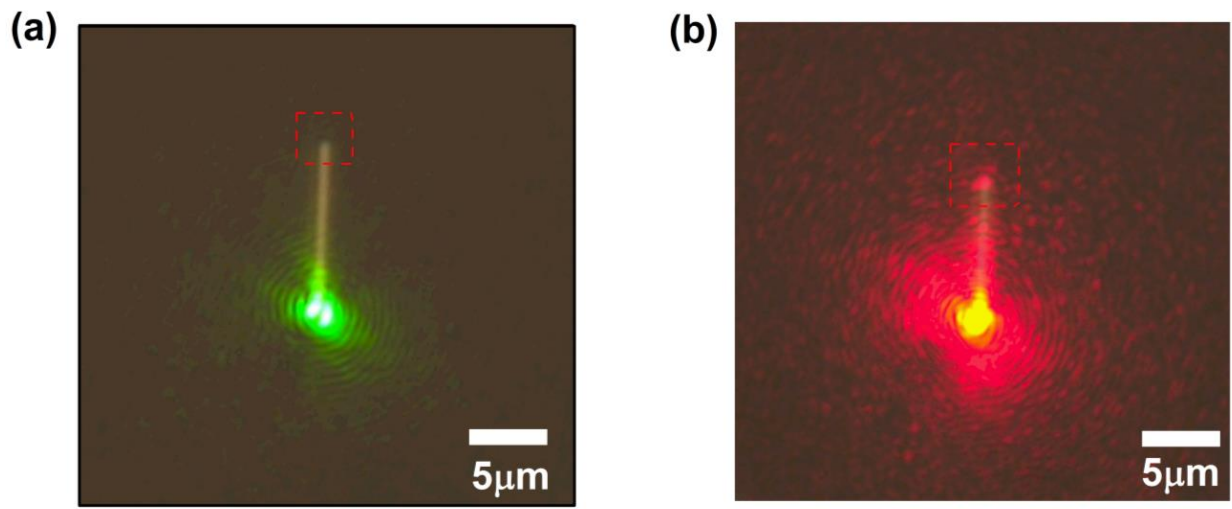

**Figure S3:** Bright field image of Wire on WS$_2$ without gold mirror in between. (a) No propagation is observed on glass coverslip because of high absorption from WS$_2$ monolayer. (b) Propagation of nanowire plasmons along the length of the nanowire when the laser used is 633 nm. With 633nm laser, the absorption is very minimal and the plasmons out-couple at the distal end of nanowire.

## 4. Fourier plane images from multiple nanowires

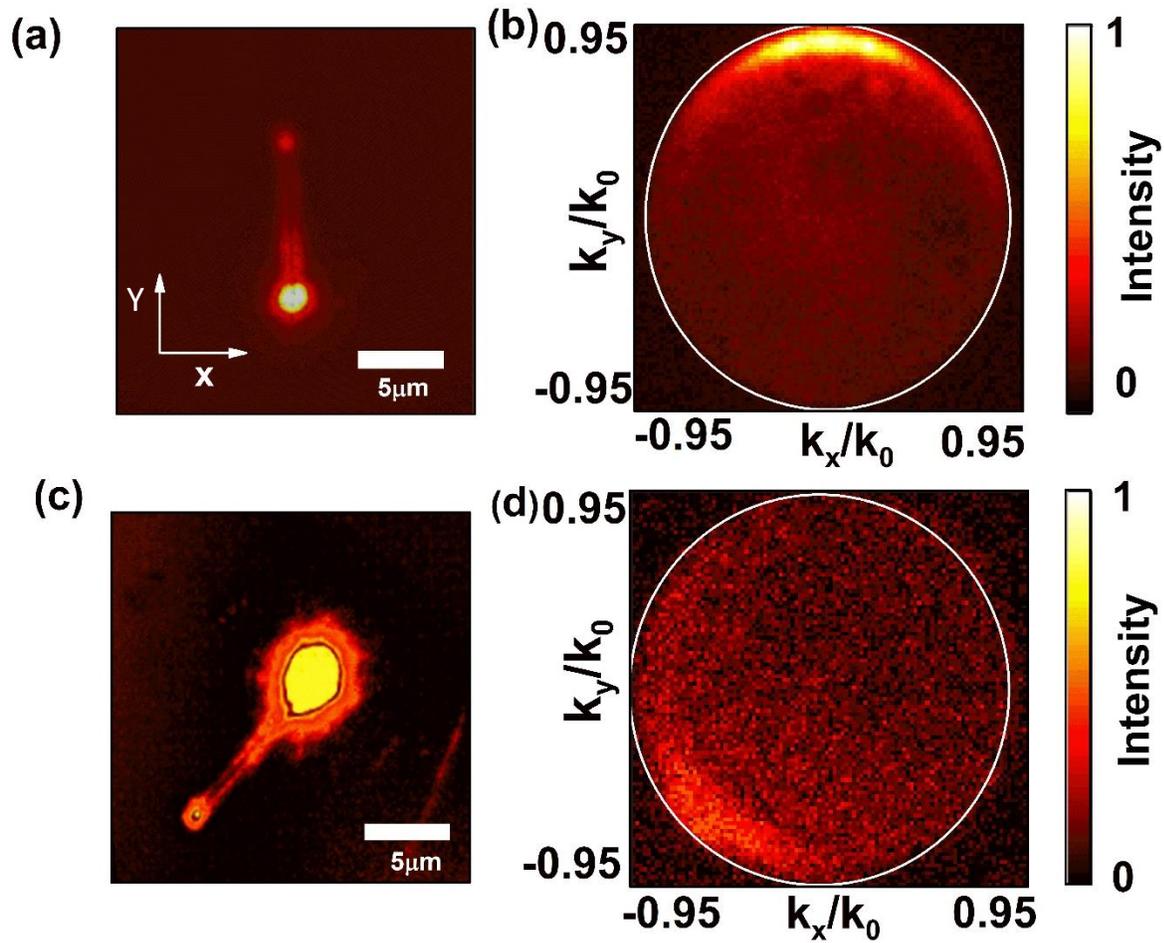

**Figure S4.** Multiple data set showing directional emission from the distal end of the nanowire. (a) and (b) are the real and Fourier plane image of a nanowire showing directional emission. (c) and (d) are the same real and corresponding Fourier plane image for the another nanowire wire showing directional emission along the length of the nanowire.

## 5. Polarization Dependent Output Spectrum deconvolution into exciton and trion peak

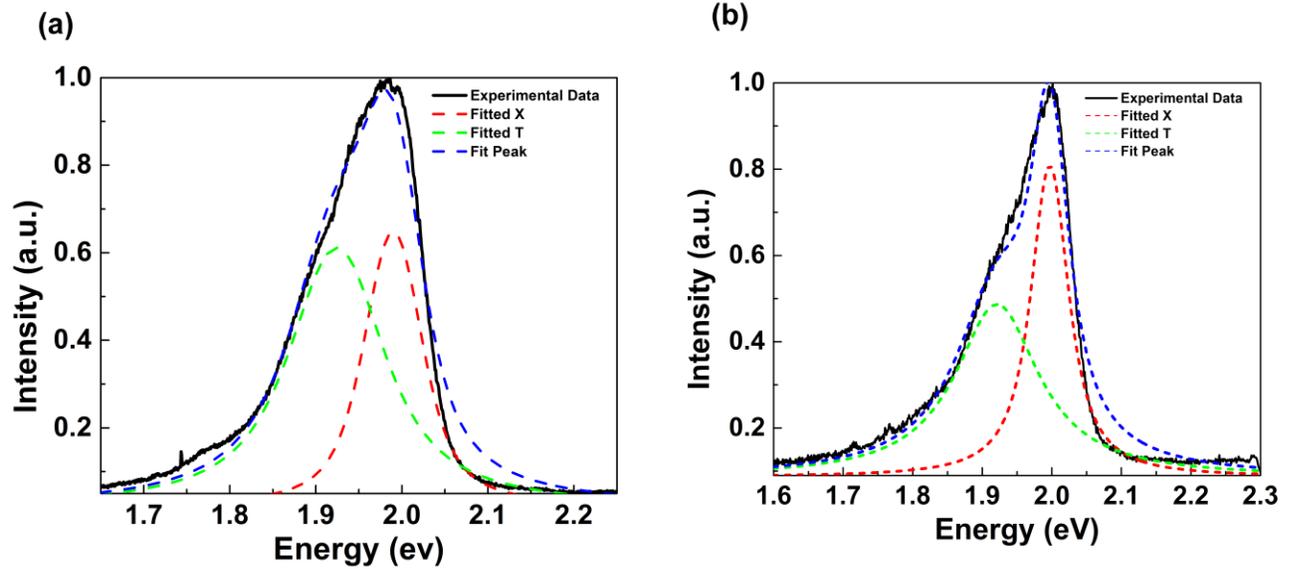

**Figure S5.** Output polarization analyzed spectrum (a) Analyzed along the long axis of the nano wire is de convoluted into exciton (Red dotted curve) and trion (green dotted curve). (b) Analyzed perpendicular to the long axis of the nano wire is de convoluted into exciton (Red dotted curve) and trion (green dotted curve).

The output analyzed spectrum in two orthogonal polarization is deconvoluted for the trion and exciton contribution. It can be seen that the trion contribution in case of polarization along the AgNW-WS$_2$-Au cavity trion contribution is higher in comparison to transverse to the AgNW-WS$_2$-Au cavity.

## 6. Mode of MIM cavity

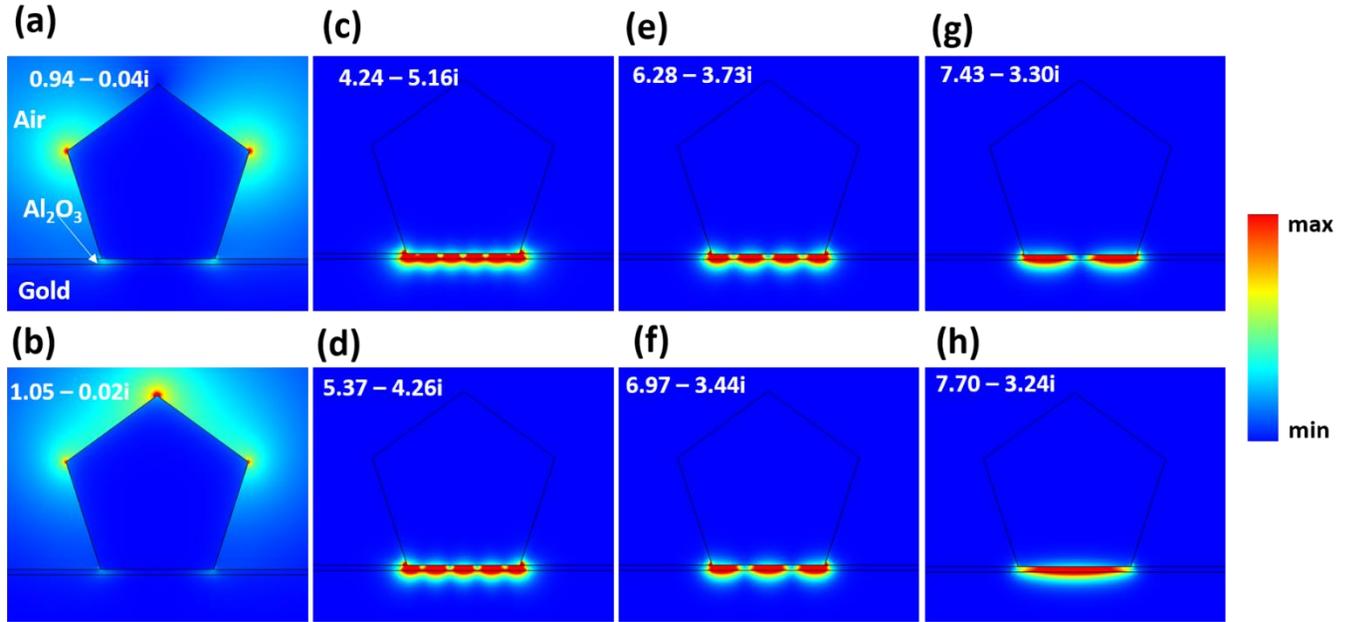

**Figure S6**: Modes of the MIM waveguide. There corresponding mode index is given. (a) and (b) are higher oreder mode. (c-h) Lossy gap plasmon mode.

To understand the role of mode of the MIM cavity we have calculated the mode of a MIM waveguide. Numerical simulation were performed using finite element methods. Ag nanowire of diameter 300 nm were placed at gold film with an intermediate layer $Al_2O_3$ of 8 nm for the mode of MIM waveguide. The length of the wire is considered infinite to calculate the possible modes due to the pentagonal symmetry of the nanowire.

For MIM waveguide we observed eight modes. The high order SPP mode (Fig. R2 (a) and (b) have larger propogation length as their imaginary part of effective mode index is small. In contrast, the six bound modes (Fig. R2 (c)-(h)) exhibit high loss owing to very high imaginary component of the mode index.

## 7. FESEM image of the NWs

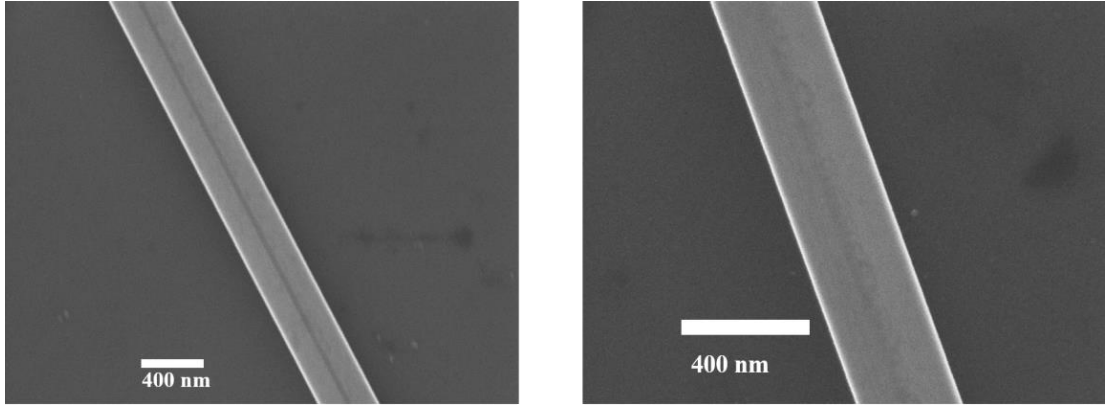

Figure S7: FESEM image of the two typical NWs